\documentclass[aps,11pt]{revtex4-1}

\newtheorem{theorem}{Theorem}

\newtheorem{lemma}{Lemma}

\usepackage{graphicx}
\usepackage{amsfonts}
\usepackage{amsmath}
\usepackage{mathrsfs}
\usepackage{eucal}
\usepackage{appendix}
\usepackage[ruled]{algorithm2e}
\usepackage{natbib}

\newcommand\bR{\mathbb{R}}

\newcommand\mF{\mathcal{F}}

\newcommand\mQ{\mathcal{Q}}

\newcommand\mA{\mathcal{A}}

\newcommand\bC{\mathbf{C}}
\newcommand\bA{\mathbf{A}}
\newcommand\bD{\mathbf{D}}

\def\o{\omega}

\newcommand\qed{$\square$}

\begin{document}



\title{Towards the full information chain theory: question difficulty}
\author{E. Perevalov}
\email[E-mail: ]{eup2@lehigh.edu}
\author{D. Grace}
\email[E-mail: ]{dpg3@lehigh.edu}
\affiliation{Lehigh University\\ Bethlehem, PA}
\date{\today}

\begin{abstract}
A general problem of optimal information acquisition for its use in decision making problems is considered. This motivates the need for developing quantitative measures of information sources' capabilities for supplying accurate information depending on the particular content of the latter.
In this article, the notion of a real valued difficulty functional for questions identified with partitions of problem parameter space is introduced and the overall form of this functional is derived that satisfies a particular system of reasonable postulates. It is found that, in an isotropic case, the resulting difficulty functional depends on a single scalar function on the parameter space that can be interpreted -- using parallels with classical thermodynamics --  as a temperature-like quantity, with the question difficulty itself playing the role of thermal energy. Quantitative relationships between difficulty functionals of different questions are also explored.
\end{abstract}

\pacs{02.50.Cw, 02.50.Le, 89.70.Cf}

\keywords{information; information theory; decision making; question difficulty; entropy}

\maketitle

\section{\label{s:intro}Introduction}
It would be almost a banality to say that information permeates all human activity: from everyday life to science and engineering. No conscious action is made without some information received and processed. In spite of its overwhelming importance, the science that has information as its main subject is still lagging somewhat behind, with classical Information Theory being the only branch that is reasonably well developed at present time. Information Theory, as we know it now, is predominantly a theory of optimal transmission which by design is unconcerned about the content of the transmitted information.

Summarizing the overall notion of information as it can be derived from experience, it would be not unreasonable to say that, at a high level of description, information can be generally characterized by its {\it quantity}, {\it accuracy} and {\it relevance}. Moreover, these three characteristics cannot in general be reduced to each other and thus can be treated as independent. Clearly, while information quantity can be studied irrespective of its possible content, the latter has to be taken into account if one desires to describe both accuracy and relevance of information in question. Additionally, these two characteristics can only be sensibly discussed relative to a particular problem which an agent attempts to solve with the help of the information. In this context, roughly speaking, accuracy of the information measures the degree of agreement between the information conveyed (by a suitable information source) to the agent and the ``true'' state of affairs. The relevance, on the other hand, would measure the degree of influence the possession of (accurate) information of a particular kind can have on the solution quality for the problem in question -- assuming the agent makes full use of it. If one looks at the classical Information Theory from this point of view, it can be said that it deals with information quantity while being explicitly indifferent to both its accuracy and relevance to any problem. In fact, it is also clear that the overall problem of information {\it transmission} (as opposed to information {\it acquisition} and {\it usage}) that the classical Information Theory addresses would in general not require any kinds of explicit reference to both accuracy and relevance of the information being transmitted.

\begin{figure}
\includegraphics{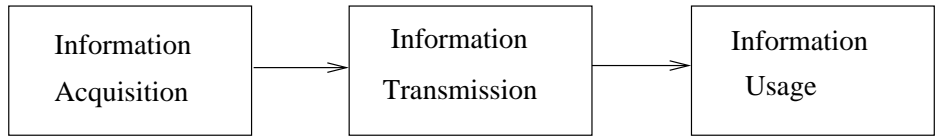}
\caption{\label{f:Ichain}Full information chain}
\end{figure}

Pursuing the point of view described in the previous paragraph, one could also look at the overall problem of information study from the angle of a {\it path} information typically takes in any conscious act of making use of it. Namely, information is first acquired (from some kind of a source) then (possibly) transmitted and, finally, used to solve some problem. One can therefore speak of the three basic links of the {\it full information chain} (see Fig.~\ref{f:Ichain}). In this picture, the classical Information Theory is the theory of the middle link. In most applications, the middle link has the convenient property of being amenable to study in isolation of the two ``end'' links of the full information chain. On the other hand, it appears that the end links have to be studied together. Indeed, the problem being solved would in general dictate the particular kind of information that needs to be acquired from an information source. On the other hand, the ``knowledge structure'' of the particular source would typical affect the potential solution quality of the given problem.

\begin{figure}
\includegraphics{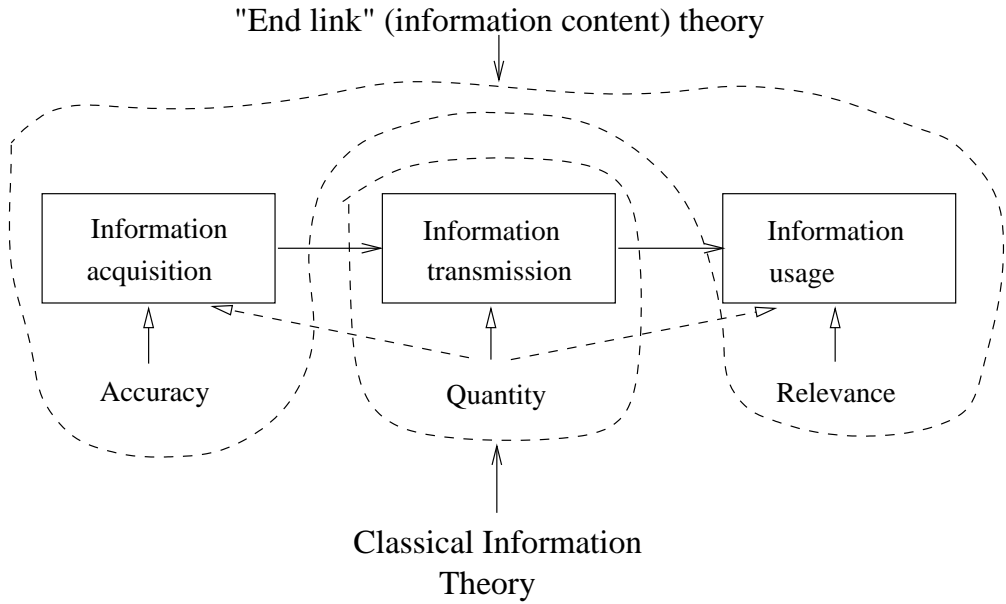}
\caption{\label{f:info-chain-char} Full information chain and information attributes.}
\end{figure}

One can now connect the two points of view (the ``information attributes'' and ``information path'' ones) to arrive at the schematic picture illustrated in Fig.~\ref{f:info-chain-char}. With information quantity being the only attribute associated with the middle link of information chain, that link should allow for a ``cleaner'' treatment which was indeed undertaken in classical Information Theory. On the other hand, the two end links of the information chain are associated with information accuracy and relevance attributes, respectively (with information quantity also playing a role), and have to be studied simultaneously which potentially complicates the analysis necessitating, in particular, a joint consideration of problems and information sources as well as modeling the ``knowledge structure'' of the latter.

This article's main purpose is to initiate systematic study of the two end links of the full information chain. More specifically, we begin with the first (information acquisition) link, with the treatment of the third link to follow soon in future publications. Since, as was mentioned earlier, the two links are closely connected, the main practical results (such as algorithms for optimal information acquisition for the particular problem) will appear in these later publications. Still more specifically, the main subject of the present article has to do with ways of soliciting the specific information from an information source and with modeling the expected accuracy of the information obtained thereby.

The discussion begins with a problem that the agent is willing to solve. To make it more specific, we assume that the problem is characterized by a well-defined real valued objective function that has to be minimized. Also, the agent is assumed to have only incomplete information about some parameters of the problem -- hence the need for additional information. Incomplete information is described self-consistently via probabilities -- therefore the initial information available to the agent is assumed to take the form of a certain probability measure on the problem parameter space which he or she looks to update by soliciting additional information from an information source. The latter process can be given the appropriate structure by enabling the agent to formulate specific requests for information, or {\it questions}, to the information source that are understood as descriptions of the specific kind of additional information the agent would want to obtain from the source. In turn, the source should be assumed to be able to provide appropriately structured {\it answers} to the agent's questions which fulfil the requests for information contained in them. The questions should in principle be selected to maximize the impact of the corresponding answers on the (appropriately measured) problem solution quality. To achieve this goal, the questions have to be selected in such a way that, on one hand, the source is able to provide accurate answers and, on the other hand, the specific information requested in the questions is relevant to the particular problem. While, as mentioned earlier, the issue of relevance will be treated in later publications, this article has accuracy as its main concern. To address the latter, note that, in general, a source would answer some questions more accurately than others. In other words, different questions appear to possess different degrees of {\it difficulty} to the source so that the more difficult questions are usually answered with lower accuracy. The agent's goal of finding questions that would be given accurate answers then becomes that of determining questions of sufficiently low difficulty to the source.

Looking at this issue from a slightly different angle, one can say that if the agent's goal is selecting questions that are preferred from the point of view of accuracy of the answers the source can provide in response to them, then it would be convenient to be able to characterize each question with a {\it real} number the value of which is directly related to the expected accuracy of the source's answer to this question. The requirement of characterizing each question with a single real number can be thought of as simply a design criterion that stems from the agent's desire to be able to completely order all questions with respect to the expected accuracy of the corresponding answers.

In this article, we introduce a class of questions that can have some relevance to the problem of interest to the agent. We then proceed to developing the real valued measure of question difficulty -- the {\it question difficulty functional}. To determine the form of the latter, it appears reasonable to begin with a set of requirements the functional has to satisfy. Such requirements can express both consistency properties and assumed symmetries of the ``knowledge structure'' of the source. The latter is understood as a quantitative description of the source's strengths and weaknesses in regards to being able to provide accurate answers to agent's requests for different types of information pertaining to the problem being solved. Depending on the degree of symmetry exhibited by the source's knowledge structure, one can obtain more or less elaborate forms of the question difficulty functional that require various geometric objects for its full description. It appears likely that there will arise a hierarchy of knowledge structure models so that a less complicated (e.g. more symmetrical) model can be obtained from a more complicated one as a limiting case (e.g. when additional symmetry becomes valid). The specific choice of a suitable model for the source would have to be made based on experience, with more complicated models replacing the simpler ones if sufficient evidence is found that the source's knowledge structure cannot be adequately described by the latter. In this article, we limit ourselves to a linear isotropic knowledge structure model postponing the discussion of more elaborate models to future publications.

\subsection{Related work}
This and the following papers can be described as an attempt to bring information theory to bear
on optimization and decision making. As is well-known, the field of Information Theory that grew out of Shannon's pioneering work on
communication theory has since had a profound impact on a number of disciplines in natural sciences
and engineering. One of fundamental advances brought by
Information Theory is the concept of entropy and mutual information that provide a natural and consistent
measure of the amount of information associated with general probability distributions. Besides a revolution in communications
which started from the demonstration that error-free transmission over imperfect channel was possible and
gave rise to the modern coding theory, the list of successful applications of these concepts includes
(but is not limited to) a simple derivation of statistical physics laws \cite{JAYNES:1957a,JAYNES:1957b}, new algorithms in computer vision \cite{viola1995}, new methods of analysis in climatology \cite{mokhov2006,verdes2005},
physiology \cite{katura2006} and neurophysiology \cite{chavez2003}. The latter were based on the concept of transfer entropy \cite{schreiber2000} which can interpreted as conditional mutual information defined for time series and that can be used to measure information transfer between different parts of complex
systems. The relatively new field of Generalized Information Theory (see e.g. \cite{klir1996})
is concerned with problems of characterizing uncertainty in frameworks that are more general than
classical probability such as Dempster-Shafer theory \cite{shafer1976}. There it was shown, for example, that
the minimal uncertainty measure satisfying consistency requirements such as general subadditivity and
additivity for combining uncertainty for independent subsystems is obtained by maximizing Shannon entropy
over all classical probability distributions consistent with the given belief
specification \cite{maeda1993,harmanec1994}.

In this paper, we use an axiomatic approach to determine the overall form of the question difficulty functional. In the context of classical Information Theory, the axiomatic approach was used, besides Shannon himself, in \cite{faddeev1956} to derive the most general form of the entropy function. Later, R\'enyi used a different set of axioms \cite{RENYI:1961} to find the one-parameter family of functions (later called R\'enyi entropies) that included standard (Shannon) entropy as a special case. The concept of structural entropy was introduced in \cite{havrda1967} and used for classification purposes. The entropy of \cite{havrda1967} (known as Havrda-Charvat entropy)
was relatively recently obtained by axiomatic means in \cite{simovici2002} where axiomatization of partition entropy was discussed on rather general grounds (see also \cite{simovici1999} for closely related work). It was shown in \cite{simovici2002} that Shannon entropy, Havrda-Charvat entropy and Gini index all obtain as particular cases of general partition entropy that satisfies a system of reasonable axioms.

 A concerted application of information methods to problems of fundamental physics that goes back to  pioneering work of Jaynes \cite{JAYNES:1957a,JAYNES:1957b} gave rise to the new field of Information Physics that produced a number of intriguing results in recent years (see \cite{caticha-rev} for a good review). The overall guiding idea behind the field is that the laws of physics are in essence the laws of inductive inference applied to physical systems. That many of known physics laws can be derived in this way has been shown for thermodynamics \cite{JAYNES:1957a}, quantum \cite{caticha11}, classical \cite{caticha07} mechanics, and (recently) relativistic quantum theory \cite{caticha12}
 The Information Physics approach shifts the emphasis in deriving laws of physics to the determination of the correct degrees of freedom and the {\it relevant information} that needs to be known for a full description of the system in question. A prototypical example would be an ideal gas in thermal equilibrium with a large bath. A full thermodynamical description of it can be obtained with the knowledge that the gas consists of molecules and that it is characterized with a definite value of the average energy per molecule -- the temperature. Once this is known, all measurable results can be obtained by an application of inductive inference rules in the form of the ME method that relies on maximization of relative entropy subject to the constraints expressing the relevant information. The ME method itself can be looked upon of as a generalization and refinement of Jaynes' original MaxEnt method.

 While the ME method addresses the problem of information {\it processing} in application to physical sciences, the issue of information {\it quantification} in Information Physics has been addressed from the point of view of order theory in \cite{knuth05,knuth07,knuth08}. This particular research direction goes back to the ideas of Cox \cite{cox1946,cox1961,cox1979} on the origins of probability. It is argued in \cite{knuth05} that order plays a more fundamental role than measure and that both probability and Shannon entropy can be derived as natural valuations on distributive lattices of assertions and questions, respectively. One of successful applications of the order-theoretic approach to fundamental physics is the recent derivation \cite{knuth-sr} of Lorentz transformations and Minkowski metric of special relativity directly from the consideration of the partial order of events in space-time.

  The approach proposed in the present and follow-up articles is rather closely related to developments in Information Physics. At a higher level, the proposed approach seeks to establish a general framework for information use in quantitative decision making in a wide variety of settings. Information Physics pursues a similar goal in application to physical sciences: it seeks to establish the general role of information in fundamental laws of nature. A bit more specifically, as mentioned earlier, while the classical Information Theory deals with information quantity, Information Physics is mostly concerned with information relevance for the given physical system. The proposed approach deals with information relevance for the problem being solved and information expected accuracy with respect to the ``true'' state. It should also be mentioned that the question difficulty functional developed in this article can be looked upon as a generalization of question relevance measure \footnote{Note that we use the term ``relevance'' for the degree to which a perfect answer to question improves the solution quality for the given problem. So relevance in our framework is defined relative to a given problem (and quantified by a number in the interval $[0,1]$). The relevance of Information Physics as defined in \cite{knuth05,knuth08} is similar to question difficulty in our approach. In fact, the relevance of \cite{knuth05} was originally called ``bearing'' in \cite{cox1979} which, in our opinion, describes the meaning of it a bit more accurately. } discussed in \cite{knuth05,knuth07,knuth08} which quantifies the degree to which a partition question resolves the central issue (the most detailed partition question). In fact, we suspect that the difficulty functional of the present article can be derived using the order-theoretic approach of \cite{knuth05}, as a valuation on the lattice of questions that's allowed to depend, besides probabilities of corresponding assertions, on suitable geometric structures on the problem parameter space. This issue is currently under investigation.

The main motivation for this and follow-up papers was the authors' desire to develop methodology for optimal use of (additional) information in decision making under uncertainty. This idea is obviously not entirely new and it has been studied and
used, for instance, in the area of statistical decision making. Applications to innovation
adoption \cite{mccardle1985,jensen1988}, fashion decisions \cite{fisher1996} and vaccine composition decisions for flu immunization \cite{kornish2008} can be mentioned in this regard. It's interesting to observe that the amount of information in these applications is typically measured simply as the number of relevant observations which can be either costless or costly, depending on the model. Some authors \cite{fischer1996}, \cite{ellison1993} introduced various models
(e.g. effective information model) for accounting for the actual, or effective, amount of
information  contained in the received observations. The common
theme of this line of work is to try to find an optimal trade-off between the amount of additional
information obtained and the suitably measured degree of achieving the original goal. Thus, for
instance, in \cite{kornish2008}, waiting longer allows the decision makers to obtain more precise
forecast of which flu virus strains are going to be predominant but leaves less time for actual
vaccine production. The main difference of the approach initiated by this paper is in that it allows to optimize not only the quantity of the acquired information but also its content and that it explicitly accounts for properties of information sources.

Explicit consideration of information sources that lies at the core of the proposed methodology is similar in spirit to analyzing and using information provided by human experts. In fact, in many practically relevant application the role of multi-purpose information sources used in the proposed approach will likely be played by human experts. In existing research literature, the problem of optimal usage of information obtained from human experts has been addressed mostly in the form of updating the decision maker's beliefs given probability assessment from multiple experts \cite{french1985,genest1986,clemen1987,clemen1999},
and, in particular, optimal combining of expert opinions, including experts with incoherent and missing outputs \cite{predd2008}.  Investigations on using and combining information of experts that partition the event differently \cite{bordley2009} and on rules of updating probabilities based on outcomes of partially similar events \cite{bordley2011} should also be mentioned in this regard. The latter investigations consider experts that provide qualitatively different information. The dependence of the quality of experts' output on the particular partition was also studied in \cite{fox2005}. In the approach developed in this and consecutive papers, the emphasis is on an explicit modeling of experts' knowledge structure and on optimizing on the particular type of information for the given expert(s) and the given decision making problem.

\subsection{Outline}
The rest of the article is organized as follows. Section~\ref{s:prelim} contains necessary preliminaries including a short discussion of information relevance. In Section~\ref{s:frame}, the adopted model of information acquisition is described and, in particular, a definition of a question to be used later is given.  Section~\ref{s:difficulty} is devoted to a discussion of the question difficulty functional -- the main topic of the present article. In particular, the main theorem establishing the overall form of the (isotropic) question difficulty functional required to satisfy certain reasonable postulates is proven. In Section~\ref{s:relations}, relationships between different questions are explored. Section~\ref{s:examples} contains simple numerical examples illustrating the results obtained earlier in the article. Finally, a conclusion summarizing the main results is given in
Section~\ref{s:conclusion}.


\section{\label{s:prelim}Preliminaries}
As mentioned in section~\ref{s:intro}, the proposed framework derives information relevance from (broadly understood) decision making problems: information is considered more relevant if it improves the corresponding decision quality to a larger extent. As the present article is devoted to information accuracy characteristic, we will consider the relevance aspect of information in detail in later publications. As far as the particular decision making problem is concerned, two aspects of it are important for our goals: the ``uncertainty'' space and the loss functional. The former supplies the description of both the initial information and the additional information provided by an information source. The latter quantifies the respective quality of the decision making problem solution and will be dealt with in later publications.

A useful picture to have in mind while reading the rest of this article is that of an agent asking questions of an information source with the latter providing answers. The source is in general not capable of providing perfectly accurate answers to the agent's questions. One could say, informally, the the source is not 100\% trustable. The source's answers can, in fact, can range from perfectly accurate to vacuous (some details are discussed in the next section). The question difficulty -- the main subject of the present paper -- is a useful construct that can help the agent predict the degree of the source's answer to a given question. As was mentioned in the Section~\ref{s:intro}, it assigns to each question a real number that gives a degree of difficulty of the question to the source and, as such, determines the degree of (expected) accuracy of the source's answer to it. The detailed characterization of properties of the source's answers and its capabilities is given in the companion paper \cite{part2}.

Let $\Omega$ be the base space that is understood as the space of possible values of input data parameters that are not known with certainty, for the problem the agent is solving. It is often referred to as a {\it parameter space}. $\Omega$ can be finite or infinite, such as a closed subset of a Euclidean space $\bR^s$. We denote by $\mF$ a suitable sigma-algebra on $\Omega$. $P$ is a probability measure on ($\Omega,\mF)$ that describes the initial state of information available to the agent and that can be modified by querying information sources. We often refer to it as a measure on $\Omega$, omitting an explicit specification of $\mF$ unless needed. In order to formalize the process of extracting information from sources we will need to describe questions. The latter task necessitates the usage of various subsets of the parameter space $\Omega$ and collections of such subsets.

A generic collection of (distinct) subsets $\bC=\{C_1,\dotsc, C_r\}$ will be called {\it inclusion-free} if for any $C_i, C_j\in \bC$ neither of the two is a proper subset of the other. A collection $\bC=\{C_1,\dotsc, C_r\}$ will be called {\it complete} if it fully covers $\Omega$, i.e. $\cup_{i=1}^r  C_i =\Omega$.

Of particular interest to us will be collections of subsets $\bC$ that are {\it partitions} of $\Omega$, meaning that all sets in $\bC$ are non-overlapping, i.e. $C_j\cap C_l=\emptyset$ for all $j\ne l$. Note that our definition of a partition differs somewhat from the standard one in that it does not require completeness which is expressed by the relation $\cup_{j=1}^r C_j=\Omega$. We refer to partitions that satisfy the completeness requirement as {\it complete partitions} and to those that do not satisfy it -- as {\it incomplete partitions}. For any partition $\bC$, we will use the notation  $\hat C\equiv \cup_{i=1}^r C_i$. Clearly, partition $\bC$ is complete if and only if $\hat C = \Omega$.

A partition $\tilde\bC$ will be called a {\it refinement} of $\bC$ if every set from $\tilde\bC$ is a subset of some set from $\bC$. In such a case, $\bC$ is a {\it coarsening} of $\tilde\bC$. Given measure $P$ on $\Omega$, we call partition $\bC_f(P)$ the {\it finest} partition of $\Omega$ associated with measure $P$ if $P(C)>0$ for all $C\in \bC_f(P)$ and there exists at least one set of zero measure in any refinement of $\bC_f(P)$. In case $\Omega$ is a closed subset of a Euclidean space and $\mF$ is a Borel algebra, it is easy to see that finest partitions do not exist if measure $P$ has continuous support or has a component with continuous support. It is also clear that if the measure $P$ has discrete support there exist many partitions of $\Omega$ that are finest for $P$.

\begin{figure}
\includegraphics[scale=0.6]{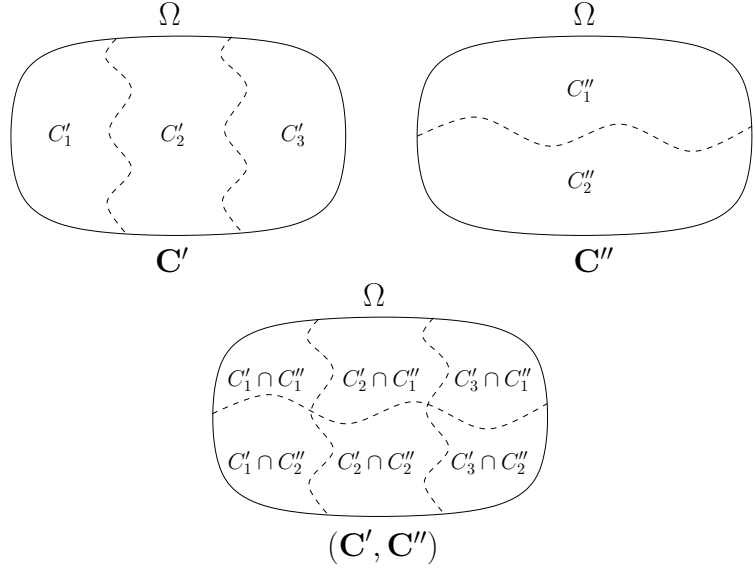}
\caption{\label{f:partitions}Two partitions of $\Omega$ and the corresponding joint partition.}
\end{figure}

Let $\bC'=\{C_1',\dotsc, C_r'\}$ and $\bC''=\{C_1'',\dotsc, C_s''\}$ be two partitions of $\Omega$. Then the partition $\bC=\bC'\cap \bC''$ is defined as the partition that consists of all sets of the form $C_i'\cap C_j''$: $\bC'\cap \bC''=\{C_1'\cap C_1'', C_1'\cap C_2'', \dotsc, C_r'\cap C_s''\}$ (see Fig.~\ref{f:partitions} for an illustration).  Obviously, some of the sets constituting partition  $\bC'\cap \bC''$ may be empty. Clearly, the partition  $\bC'\cap \bC''$ is a refinement of both $\bC'$ and $\bC''$.

\begin{figure}
\includegraphics[scale=0.6]{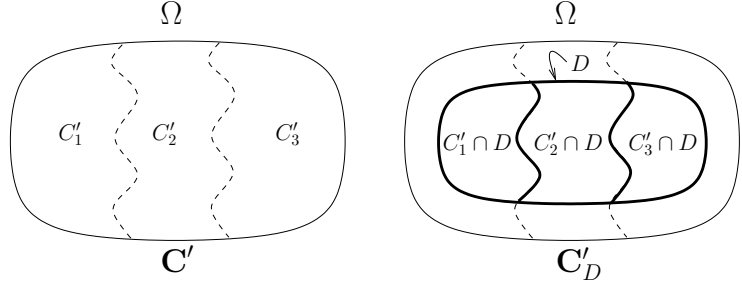}
\caption{\label{f:partition_induced}Partition $\bC'_D$ of set $D\subset\Omega$ induced by a partition $\bC'$ of $\Omega$.}
\end{figure}

If $D\subset \Omega$ is a subset of $\Omega$ and $\bC'=\{C_1',\dotsc, C_r'\}$ is a partition of $\Omega$, the partition $\bC'_D=\{D\cap C_1',\dotsc, D\cap C_r'\}$ of $D$ will be called the partition of $D$ {\it induced} by the the partition $\bC'$ of $\Omega$ (see Fig.~\ref{f:partition_induced}).

Let $C\in \mF$ be a measurable subset of $\Omega$. We denote by $P_C$ the conditional measure on $\Omega$ defined as
\begin{equation}
P_C(D)=\frac{P(D\cap C)}{P(C)},
\label{eq:cond-measure}
\end{equation}
for arbitrary $D\in \mF$.

For an arbitrary complete partition $\bC=\{C_1,\dotsc, C_r\}$, it is straightforward to show that the following decomposition of the measure $P$ into the corresponding conditional measures
\begin{equation}
P=\sum_{j=1}^r P(C_j)P_{C_j}
\label{eq:P-decomp}
\end{equation}
is valid.

\section{\label{s:frame}Information acquisition}
In the model of information acquisition we adopt, an information source is assumed to be capable of providing answers to questions. The accuracy of the source's answers generally varies with the question's
nature: a given source can answer some questions more accurately than others. This difference in answer accuracy reflects the information source's {\it knowledge structure} which we model by assigning a real number to each possible question a source can receive. This number is naturally termed the {\it question difficulty}, the meaning of it being that the source can provide less accurate answers to questions of higher difficulty. Just like a question can be characterized with its difficulty, an answer to it can be characterized with its {\it depth}, with more accurate answers to the same question having a higher depth value. We address the question difficulty here, postponing a discussion of answer depth to a follow-up paper \cite{part2}. First, we need to define what a question is.

\subsection{Questions}
A question is a request for new information over what is already known. The latter is represented by the probability measure $P$ on the parameter space $\Omega$ and is assumed to be common knowledge. One can think of $\Omega$ as a set of all possible states of the system in question. Let us assume, for simplicity, that $\Omega$ is a finite discrete set of cardinality $N$. This assumption is not going to be limiting as one can always apply a suitable discretization to a continuous parameter space $\Omega$. Since there is typically no order imposed on the elements $\o$ of $\Omega$, they form an {\it antichain}. Then, as shown in \cite{knuth05}, the space of {\it assertions} over the elements of $\Omega$ can be represented as a {\it Boolean lattice} \footnote{Here, a lattice is a partially ordered set $J$ such that any pair of elements of $J$ has a unique lowest upper bound (join) and a unique highest lower bound (meet) with respect to some binary ordering relation generically denoted by $\le$.} $\mA_N$ where the ordering relation $\le$ is identified with logical implication ($x\le y$ if and only if $x$ implies $y$) and the lattice operations of join and meet are identified with logical disjunction and conjunction, respectively.

A question was originally defined by Cox \cite{cox1979} as a set of logical assertions that answer it fully.
Using this definition, one can construct \cite{knuth05} a lattice of questions $\mQ_N$ from the Boolean lattice $\mA_N$ of logical assertions. The set of {\it ideal} questions -- which, if ordered by set inclusion, forms a lattice isomorphic to $\mA_N$ -- is built from {\it down-sets} \footnote{A down-set of any subset $K$ of a partially ordered set $J$ is the set of all elements $y$ of $J$ such that $y\le x$ for some element $x$ of $K$.} \cite{davey2002} of the elements of $\mA_N$. More generally, the set of all questions comprising  $\mQ_N$ can be formed by taking all distinct down-sets of subsets of $\mA_N$ (see \cite{knuth05} for details). As was mentioned above, $\mQ_N$ naturally becomes a lattice where the ordering relation $\le$ is identified with ``answering'' (i.e. $x\le y$ if and only if any full answer to $x$ also fully answers $y$).

It is easy to see from the construction outlined above that, in case when the system states form an antichain $\Omega$, questions can be identified with all possible {\it inclusion-free} collections of subsets of $\Omega$. In particular, ideal question are identified with individual subsets of $\Omega$. The {\it central issue} was defined in \cite{knuth05} as the down-set of the union of all atoms (logical assertions corresponding to individual elements of $\Omega$) of $\mA_N$. All questions that lie above the central issue in $\mQ_N$ (i.e. such questions that are fully answered by any logical assertion fully answering the central issue) were called {\it real questions} by Cox and also in \cite{knuth05}. The set of real questions form a sublattice of $\mQ_N$, with the smallest real question (the bottom of the sublattice) being the central issue. It can easily be seen that, in terms of collections of subsets of $\Omega$, all real questions are {\it complete collections}, with the cental issue being identified with the finest partition of $\Omega$. Note that only one ideal question -- the one corresponding to the whole $\Omega$ -- belongs to the set of real questions. In the following, we will concentrate on  questions corresponding to both complete and incomplete partitions of $\Omega$ which we refer to as {\it complete partition questions} and {\it incomplete partition questions}, respectively. All former -- called simply {\it partition questions} by Knuth in \cite{knuth05,knuth08} -- are real questions \footnote{Note that not all real questions correspond to complete partitions.}, and the latter belong to the set of {\it vain questions} -- the complement of the set of real questions inside $\mQ_N$ -- as defined by Cox. Note that incomplete partition questions will play a mostly auxiliary role, with all the practically important examples being of the complete partition -- and hence real -- variety. Thus we adopt the following definition.

{\bf Definition:} A question is a partition $\bC=\{C_1, C_2,\dotsc, C_r\}$ where $C_j$, $j=1,\dotsc, r$ are subsets of $\Omega$ such that $C_i\cap C_j=\emptyset$ for $i\ne j$ and $\cup_{j=1}^r C_j\subseteq \Omega$.

For any partition $\bC$ we denote the union of all subsets in $\bC$ by  $\hat C$:
$\hat C\equiv \cup_{j=1}^r C_j$. Thus, for any complete partition $\bC$, $\hat C = \Omega$.

In everyday terms, a complete partition question (which is always real in Cox's terminology)
can be interpreted as a traditional multiple-choice question (e.g. {\it ``Is the apple red, green or yellow?''}. Note that in most applications one would typically be dealing with questions that are above (actually, quite often well above) the central issue in the $\mQ_N$ lattice. One can say that such questions do not request the fullest possible information about the system (problem) in question. For example, the question just mentioned does not request any information about the apple size. Observe that our definition of a question is close to that proposed in \cite{caticha04} since any (complete) partition of the base space $\bC$ induces the corresponding probability distribution $P(\bC)=(P(C_1),\dotsc, P(C_r))$ and thus the question $\bC$ can be interpreted as a probability distribution which, in turn -- since it represents the agent's incomplete information -- can be thought of a request for missing information.

 Incomplete partition questions are typically more difficult to verbalize and interpret.
 Knuth \cite{knuth05} refers to them as auxiliary constructs similar to negative numbers that are not directly used for counting but can be very useful to assist it. We will discuss this issue in more details later in the paper, when the notion of an answer is introduced.

 Since the only types of questions we consider in this and follow-up articles are partition questions (complete or incomplete), we will often use the terms ``partition'' and ``question'' interchangeably.
 In particular, we will often say ``complete question'' and ``incomplete question'' to refer to a complete partition question and incomplete partition question, respectively. We will also refer to incomplete partition questions with partitions consisting of a single subset as {\it ideal questions} making use of the established terminology.

\subsection{Answers}
Recall that Cox \cite{cox1979} defined a question as a collection of logical assertions that provide a (full) answer for it. Respectively, given a question an answer would have to be identified with any of such logical assertions. Note, in particular, that in this picture answers that provide more information than the question requests are perfectly admissible. For example, if the question reads {\it ``Is the apple green or not?''}, the answer stating that the apple is red would be a valid answer to the question. In this and following articles, we take a somewhat different view in which an answer to the given question provide {\it no more} than the requested information -- as opposed to {\it no less} in Cox's picture. Informally speaking, a typical answer to the question mentioned above would be {\it ``The apple is more likely to be green than not''}, with different degrees of confidence in the stated assertion possible. In particular, one of key points in our approach lies in explicit consideration of imperfect answers,  coupled with quantification of the degree of accuracy of such answers.

While a detailed discussion of answers and their accuracy related properties will be given in the companion paper \cite{part2}, we present some of it here since it is needed to clarify the notion of questions and, especially, incomplete -- including ideal -- questions. We begin with perfect answers.

{\bf Definition:} Given a question $\bC=\{C_1,\dotsc, C_r\}$, the {\it perfect answer} $V^*(\bC)$ is a message that takes one of the values in the set $\{s_1,\dotsc, s_r\}$ such that a reception of the value $s_k$ of the message has the effect of modifying the original probability measure $P$ on $\Omega$ to the measure $P^k$ such that $P^k(C_j)=\delta_{kj}$ and $P^j_{C_j}=P_{C_j}$ for all subsets $C_j$ in $\bC$.

Informally speaking, a perfect answer to $\bC$ completely resolves the uncertainty associated with the partition $\bC$, i.e. places a random outcome $\o$ in one of the subsets in $\bC$ with certainty but otherwise does no more (since it leaves the conditional measures $P_{C_j}$ unchanged). Let us now generalize the definition of an answer to include possibilities of source errors. We assume, without loss of generality, that $P(C_j)>0$ for all subsets $C_j$ in the partition $\bC$.

{\bf Definition:} An answer to the question $\bC=\{C_1,\dotsc, C_r\}$ is a message $V(\bC)$ that takes values in the set $\{s_1,s_2,\dotsc, s_m\}$ such that a reception of the value $s_k$ of the message updates the initial measure $P$ on $\Omega$ to the measure $P^k$ such that either $P^k(C_j)=0$ or $P^k_{C_j}=P_{C_j}$ for all $k=1,\dotsc m$ and all $j=1,\dotsc, r$.

The difference between these two definitions is in that the condition $P^k(C_j)=\delta_{kj}$  that makes an answer perfect is not required in the general case. Moreover, the number of values a general answer can take does not have to be equal to the number of subsets in the partition $\bC$. Informally speaking, this describes the possibility of the source conveying different ``gradations of belief'' in various assertions. For instance, if the question is {\it ``Is it an apple, a pear or a peach?''}, the source could give an answer of the sort {\it ``Almost surely an apple.''} or, {\it ``More than likely an apple.''} or something of the kind. Alternatively, if the answer is not required or assumed to be perfect, the source could give just the traditional ``assertion-type'' answers, but the accuracy of them could be less than 100\%, For example, if, according to the initial measure $P$, apple, pear and peach are equally likely, the probability that the fruit is really an apple given the answer {\it ``Apple.''} by the source could be found to be equal to 0.6. On the contrary, for a perfect answer to this question, the probability of the fruit being an apple following an answer {\it ``Apple.''} by the source has to be equal to 1 by definition.

It is straightforward to show for $V(\bC)$ to be an answer to a complete question $\bC$ according to this definition, it is necessary and sufficient for the updated measures $P^k$, $k=1,\dotsc, m$, to take the form
\begin{equation}
P^k=\sum_{j=1}^r p_{kj}P_{C_j},
\label{eq:Pk-mc}
\end{equation}
where $p_{kj}$, $k=1,\dotsc, m$, $j=1,\dotsc, r$ are nonnegative coefficients such that $\sum_{j=1}^r p_{kj}=1$ for $k=1,\dotsc, m$. If the corresponding answer is perfect, $p_{kj}=\delta_{kj}$. In the following we will denote the probability that by answer $V(\bC)$ takes the value $s_k$ -- by $v_k$.

If question $\bC$ is complete and $V(\bC)$ is a corresponding answer, it is reasonable to assume that $V(\bC)$ does not change the original measure $P$ on average, or, in other words, that the original measure $P$ is a ``valid'' one that only gets ``refined'' by an answer to $\bC$. Formally speaking, this assumption means that
\begin{equation}
\sum_{k=1}^m v_k P^k = P,
\label{eq:cons-mc}
\end{equation}
from which it follows, in particular, that if the answer is perfect, then $v_j=P(C_j)$. We refer to (\ref{eq:cons-mc}) as the {\it consistency with prior} condition for the answer $V(\bC)$.

Since incomplete and, in particular, ideal questions will be made use of -- if in an auxiliary sense only -- in the next section, we find it appropriate to discuss their interpretation in some more detail. In the following, we use the term {\it correct answer} for a question (complete or incomplete) to denote a full description of one of the subsets in the corresponding partitions. Thus, a question of the form $\bC=\{C_1,\dotsc, C_r\}$ has a total of $r$ correct answers. A correct answer can be thought of as particular ``state of the world'' which becomes known only after the particular random outcome $\o$ from the parameter space $\Omega$ is observed. A correct answer to a question is in general uncertain at the time the agent need to make a decision, unless a source capable of producing a perfect answer to the corresponding question is available. Without excessive terminology abuse, one can say that a perfect answer and a correct answer relate to each other as a random variable and its particular realization.
For example, if the question is {\it ``Is this fruit an apple, a pear, or a peach?''} then {\it ``Apple''}, {\it ``Pear''} and {\it ``Peach''} are the possible correct answers, and a perfect answer is a message that can take three values such that a reception of each value of the message identifies the correct answer with certainty.

This implies, for instance, that any ideal question has a unique correct answer.  We interpret an ideal question as a real question {\it conditioned on some correct answer value}. For example, if a source is shown an apple then an ideal question can sound like {\it ``What kind of fruit is it?''}. Moreover, if an apple, a pear and a peach are the only three possible kinds of fruit that ``exist'' in $\Omega$, then a question {\it ``What kind of fruit is it?''} is clearly a real question that always (in any ``true state of the world'') has a correct answer. But, for a {\it given} ``state of the world'' (correct answer) the same question can be considered an ideal question. One could say that a source capable of a perfect answer actually can identify which ideal question (out of the three possible) is being asked. The agent, on the other hand, at the time of formulating a question can only consider it as a real question. One therefore can say that an ideal question represents a certain ``side'' or ``aspect'' or a real question. Note in this regard that the same ideal question (described by the subset of $\Omega$ corresponding to an apple) can play a role of an ``aspect'' of different real questions: for instance, {\it ``What kind of fruit is it?''} and {\it ``Is it an apple or not?''}. As far as answers to ideal questions are concerned, we interpret them as ``conditioned'' answers to corresponding real questions.

Let $\bC=\{C_1,\dotsc, C_r\}$ be a real partition question and let $C_l$  be a particular ideal question.
Suppose $V(\bC)$ is some answer to $\bC$ such that $v_k=\Pr(V(\bC)=s_k)$, $k=1,\dotsc, m$. Let us denote by $q_k^{(l)}$ the probabilities of different value of the corresponding answers to $C_l$. Using the definition of an ideal question as a ``conditioned'' real question and using the Bayes' rule we obtain
\begin{equation*}
 q_k^{(l)}=\Pr(V(\bC)=s_k|\o\in C_l)=\frac{p_{kl}v_k}{P(C_l)}.
\end{equation*}
In particular, if the answer $V(\bC)$ is perfect, $p_{kl}=\delta_{kl}$, and, as mentioned earlier, $v_k=P(C_k)$. Therefore, in this case, $q_k^{(l)}=\delta_{kl}$, or, in words, any ideal question $C_l$, $l=1,\dotsc, r$, will have just one answer value $s_l$ which identifies the corresponding (unique) correct answer. We see that, in this case (and only in this case), an ideal question receives just a single answer value from the source. In some sense, this is ``not even a question''. But this is perfectly logical if the source is capable of providing a perfect answer to the corresponding real question: it is not ``really a question'' for the source since the source is good enough to be able to perfectly distinguish between all $r$ subsets. To see one more illustrative example as to what ideal questions are (in our interpretation), consider an instructor asking a student a multiple-choice (real) question: {\it ``Is it an apple, a pear or a peach?''}. In this situation, while both know what the real question is, only the instructor (by virtue of knowing what the correct answer is) knows which ideal question is being asked.
For instance, the ideal question corresponding to an apple, (i.e. the ideal question a correct answer to which is {\it ``Apple''}) ``exists'' only when an apple is shown to the student by the instructor. In particular, all characteristics (such as probabilities of various answers $q_k^{(l)}$) of the student's answer to this ideal question are ``oblivious'' to what the students might say when the fruit shown to him is not an apple.

This interpretation, in particular, allows us to obviate the need for using the logical absurdity element as a possible answer to ideal (and other incomplete) questions. Indeed, as explained above, an ideal question in this picture would have either a single answer (in the case when the source is capable of perfect answers to the real question one ``aspect'' of which is represented by the ideal question) or multiple answers some of which do not coincide with the ideal question's correct answer.

\section{\label{s:difficulty}Question difficulty}
We finally arrive to the point where we can introduce the notion of question difficulty. As was discussed in Section~\ref{s:intro}, the main goal of introducing the question difficulty concept is to be able to predict the expected accuracy of the given information source's answers to any questions of the form described in the previous section. By its design, the question difficulty should be a real number that is assigned to any question and that depends, besides the question itself, on the state of original information available and, possibly, on some parameters characterizing the source's knowledge structure which has be described by some construct related to the base space $\Omega$. Thus, in general, the question difficulty is a real-valued functional on $\Omega$ which we denote by $G(\Omega,\bC,P)$.

Our goal in this section is to derive a general form of this functional and -- along the way -- establish the set of parameters it can depend upon. As was discussed earlier, we proceed by formulating a set of requirements -- of both consistency and symmetry variety -- that the sought for functional would need to satisfy. As far as symmetry requirements are concerned, in this article, we consider {\it isotropic} models only, postponing the discussion of more general ones to future publications. We refer to all such requirement as postulates.

As has been mentioned earlier, in the model adopted here, incomplete questions are to be understood as auxiliary constructions, while complete questions have a clear meaning. For an incomplete question $C\subset \Omega$, the difficulty functional $G(\Omega, C, P)$ can be thought of as {\it conditional} difficulty of any complete question $\bC$ containing the subset $C$ given that the random outcome $\o$ is in $C$. For example, if the subset $C_1$ represents apple, $C_2$ -- pear and $C_3$ - peach so that $C_1\cup C_2\cup C_3=\Omega$, then $G(\Omega, C_1, P)$ can be interpreted as the difficulty of the question {\it ``Is it an apple, a pear, or a peach?''}, or, equivalently {\it ``What kind of fruit is it?''} (since the source knows that the possible types are apple, pear and peach), provided the correct answer is {\it ``Apple''}.

 One reasonable and almost obvious requirement that can be imposed on the question difficulty functional $G(\Omega, \bC, P)$ is that of {\it certainty}, i.e. the difficulty of a question should vanish if there is no new knowledge to acquire given the original state of it. Formally speaking, $G(\Omega,\bC,P)=0$ whenever $P(C_j)=1$ for some value of the index $j$. One can say that in this case the question is already answered at the time of its formulation. These are questions of the kind {\it ``Is this red apple red, green or yellow?''}. Thus we obtain

 {\bf Postulate Q1} ({\it Certainty}). Suppose $\bC=\{C_1,\dotsc, C_r\}$ and $P(C_j)=1$ for some value of~$j$. Then $G(\Omega,\bC,P)=0$.

 The second postulate we propose requires that the question difficulty functional be continuous in all its arguments (which are yet to be determined).

 {\bf Postulate Q2} ({\it Continuity}). $G(\Omega, \bC, P)$ is a continuous function of all its arguments.

 The next postulate states that, for incomplete questions that are not ideal, i.e. include several subsets, the difficulty is additive: the overall difficulty of the question is the sum of the difficulty of the ideal component and the difficulty of the complete questions that results when the ideal question was answered correctly. Formally, we obtain the following.

 {\bf Postulate Q3} ({\it Incomplete question decomposition}). Let $\bC=\{C_1,\dotsc, C_r\}$ be an incomplete question.
 Then $$G(\Omega, \bC, P)=G(\Omega, \hat C, P)+G(\hat C, \bC, P_{\hat C}). $$

 This postulate describes the difficulty of questions of the sort {\it ``What kind of fruit is it and is it red, green or yellow?''}, given that the correct answer to the first part of the question is {\it ``Apple''}. It states that the difficulty of the overall question is additive: it is equal to the sum of difficulties of two questions: {\it ``What kind of fruit is it?''} (conditioned on {\it ``Apple being the correct answer''}) and {\it Is this apple red, green or yellow?''}. Note that if the question $\bC$ is complete, the first term on the right-hand side of the statement of Postulate Q3 vanishes by Postulate Q1 and the statement of Postulate Q3 reduces to a trivial identity (since in this case $\hat C=\Omega$).

 The next postulate states the mean value property of incomplete questions: the difficulty of the question $\bC\cup \bC'$ obtained by taking the union of two incomplete non-overlapping partitions $\bC$ and $\bC'$ is equal to the arithmetic mean value of the difficulties of the constituents questions with respect to the original measure $P$.

 {\bf Postulate Q4} ({\it Mean value}). Let $\bC$ and $\bC'$ be two incomplete questions such that $\hat C \cap \hat C'=\emptyset$. Then
 $$G(\Omega, \bC\cup \bC', P)= \frac{P(\hat C)G(\Omega, \bC, P)+P(\hat C')G(\Omega, \bC', P)}{P(\hat C \cup \hat C')}. $$

 This postulate can be interpreted as follows. Let $\bC$ and $\bC'$ each consist of a single subset: $\bC=\{C\}$ and $\bC'=\{C'\}$ for $C\subset \Omega$, $C'\subset \Omega$. Assume also that $C\cup C'=\Omega$, so that $\{C,C'\}$ is a complete question. Then the statement of Postulate Q4 would read
 \begin{equation}
 G(\Omega, \{C,C'\}, P)=P(C)G(\Omega, C, P)+P(C')G(\Omega, C', P),
 \label{eq:pq4}
 \end{equation}
 which is consistent with the interpretation of the difficulty $G(\Omega, C, P)$ of an incomplete question as that of a complete question containing $C$ as one of options conditioned on $C$ being true (i.e. conditioned on $\o\in C$). For instance, let $C$ represent an apple and $C'$ a pear and assume the these are the only two possible kind of fruit. Then expression (\ref{eq:pq4}) states that the difficulty of the question {\it ``What kind of fruit is it?''} is equal to the average of difficulty of the same question over all possible correct answers. From this point of view, Postulate Q4 sounds rather natural and generic. But the real meaning of Postulate Q4 is in that it states that the conditional difficulties are independent of the number and measures of other options (subsets). Postulate Q4 assigns the same conditional difficulty $G(\Omega, C, P)$ to the subset $C\subset \Omega$ regardless of the complete partition it is a member of. For instance, if $C\subset \Omega$ represents an apple then, given that the fruit is really an apple, the difficulty of the question {\it ``Is it an apple or not?''} would be the same as that of {\it ``What kind of a fruit is it?''} even if the number of possible choices (kinds of fruit) is large. It is easy to see that this is, while not unreasonable, still may be a rather strong assumption which may not be true for realistic information sources (especially human experts). Postulate Q4 can be thought of as an expression of {\it linearity} of the difficulty functional and it can be expected to be relaxed or modified in more general models.

 To state the next postulate we need to introduce a new notion. We say that the parameter space $\Omega$ is {\it homogeneous} if the question difficulty functional depends only on its subset measures for any question $\bC$ in $\Omega$: $G(\Omega, \bC, P)=f(P(\bC))$ where $P(\bC)$ stands for the vector $(P(C_1),\dotsc, P(C_r))$. More generally, we say that a subset $D\subseteq \Omega$ is {\it homogeneous} if
 $G(D, \bC, P_D)=f(P_D(\bC))$ as long as $\hat C\subset D$. In particular, any atom (minimal set) of the sigma-algebra $\mF$ is homogeneous. Postulate 5 then states that the difficulty of an incomplete question does not depend on how it is approached (directly or in stages) as long as all the intermediate questions lie inside a homogeneous subset of the parameter space.

 {\bf Postulate Q5} ({\it Homogeneous incomplete sequentiality}). Let $D\subset \Omega$ be a homogeneous subset of the parameter space and let $\bC$ be a question such that $\hat C\subset D$. Then
 $$G(\Omega, \bC, P)=G(\Omega, D, P) + G(D,\bC,P_D). $$

 To get a little more ``feel'' for this postulate think of a question asking to identify a certain animal species. The gradual approach to such a question would involve asking intermediate questions about the class the animal belongs to, order, suborder, superfamily, family, and, finally, the species itself. In case the original question is of ``harder than average'' variety it would be easier to answer the question in stages compared to answering it right away. On the other hand, if the original question is an easy one (easier than other similar questions) it can be easier to answer it without resorting to the intermediate ``guiding'' questions. A good example of the latter would be a question about a domestic cat that an average person would be able to answer easily and correctly whereas the ``guiding'' questions about class, order etc. would likely present some difficulty. Respectively, if all such questions are equally hard (for the same measure) then it would make sense to believe that the intermediate ``guiding'' questions would not change the difficulty of the original question just like the postulate states.

 Note also that Postulate Q5 looks very similar to Postulate Q3 which does not require any homogeneity to be valid. The main difference is that the second term in the right-hand side of the main statement of Postulate Q5 involves a difficulty of an {\it incomplete} question (since $\hat C\ne D$) whereas the corresponding term in the statement of Postulate Q3 is required to describe a difficulty of a complete question with the base space $\hat C$.

 Finally, it certainly makes sense to require that if $D\subset\Omega$ is homogeneous and $C\subset D$  then $f(P_D(C))=G(D, C, P_D)$ should be a decreasing function of its argument $P_D(C)$. Indeed, an incomplete question about something ``rare'' should be more difficult. We thus obtain Postulate 6.

 {\bf Postulate Q6} ({\it Homogeneous incomplete monotonicity}). Suppose $D\subset\Omega$ is homogeneous and $C\subset D$. Then $f(P_D(C))=G(D, C, P_D)$ is a decreasing function of its argument $P_D(C)$.

 In order to get still more insight into the proposed set of postulates for the question difficulty functional consider the following alternative postulate.

 {\bf Postulate Q3$'$} ({\it Complete question sequentiality}). Let $\bC=\{C_1,\dotsc, C_r\}$ be a complete question and let $\tilde \bC$ be its refinement. Then
 $$G(\Omega,\tilde \bC, P)=G(\Omega,\bC,P)+\sum_{j=1}^r P(C_j)G(C_j,\tilde \bC_{C_j},P_{C_j}).$$

 Postulate Q3$'$ states that if a complete question is made more detailed the difficulty of the resulting question can be obtained as a sum of the difficulty of the original question and the average (with respect to the measure $P$) of difficulties of conditional detalizations. For instance if the original question was {\it ``Is it an apple or a pear?''} and the detalization sounds like {\it ``Is it an apple or a pear and is its color red, green or yellow?''} then Postulate 3' says that the difficulty of the detailed question is equal to the difficulty of the original question plus the average of difficulties of questions {\it ``Is this apple red, green or yellow?''} and the question {\it ``Is this pear red, green or yellow?''}. This postulate may seem to be somewhat more reasonable and grounded in experience compared to, for instance, the {\it Mean value} postulate. It turns out though that Postulate Q3$'$ is implied by Postulate Q3 and Postulate Q4 as the following lemma shows.

 \begin{lemma}
 Suppose Postulate Q3 and Postulate Q4 hold. Then Postulate Q3$'$ holds as well.
 \end{lemma}

 {\bf Proof:} Let $\tilde \bC$ be a refinement of $\bC=\{C_1,\dotsc, C_r\}$. Then we can write
 \begin{align*}
G(\Omega,\tilde \bC, P) &\overset{(a)}{=} \sum_{j=1}^r P(C_j)G(\Omega, \tilde\bC_{C_j}, P)\\
&\overset{(b)}{=} \sum_{j=1}^r P(C_j)(G(\Omega, C_j, P)+G(C_j, \tilde\bC_{C_j}, P_{C_j}))\\
&= \sum_{j=1}^r P(C_j)G(\Omega, C_j, P)+\sum_{j=1}^r P(C_j)G(C_j, \tilde\bC_{C_j}, P_{C_j})\\
&\overset{(c)}{=} G(\Omega, \bC,P)+ \sum_{j=1}^r P(C_j)G(C_j, \tilde\bC_{C_j}, P_{C_j}),
\end{align*}
where (a) follows from the Postulate Q4 since $\bC=\cup_{j=1}^r \tilde\bC_{C_j}$, (b) follows from Postulate Q3 and (c) follows from Postulate Q4. \qed

Thus we see that Postulates Q3 and Q4 can be regarded as a somewhat stronger version of the {\it complete question sequentiality} property expressed by Postulate Q3$'$.

If we now demand that Postulates Q1 through Q6 hold for the question difficulty functional $G(\Omega, \bC, P)$, the question for us is what form this functional can possibly take. The answer is given in the following theorem.

\begin{theorem}
Let the function $G(\Omega, \bC, P)$ where $\bC=\{C_1,\dotsc, C_r\}$ satisfy Postulates Q1 through Q6. Then it has the form
$$G(\Omega, \bC, P)=\frac{\sum_{j=1}^r u(C_j)P(C_j)\log \frac{1}{P(C_j)}}{\sum_{j=1}^r P(C_j)},$$
where $u(C_j)=\frac{\int_{C_j}u(\o)\,dP(\o)}{P(C_j)}$ and $u$: $\Omega\rightarrow \bR$  is an integrable nonnegative function on the parameter space $\Omega$.
\label{th:G}
\end{theorem}

{\bf Proof:} Let $\bA=\{A_1,\dotsc, A_M\}$ be a (complete and sufficiently fine) partition of $\Omega$. We can assume, without loss of generality, that the sigma-algebra $\mF$ on $\Omega$ is comprised of all unions of sets in $\bA$.

Let $D\subset \Omega$ be a homogeneous subset of the parameter space and let $C\subset D$ be an ideal question lying inside of $D$. Furthermore, let $C'\subset C$ be another question inside of $C$. Then, according to Postulate Q5,
\begin{equation}
G(\Omega,C,P)=G(\Omega,D,P)+G(D,C,P_D),
\label{eq:homDC}
\end{equation}
and, since $C$ is homogeneous as well,
\begin{equation}
G(D, C',P_D)=G(D,C,P_D)+G(C,C',P_C).
\label{eq:homCC'}
\end{equation}
Using the form of $G(\cdot)$ for homogeneous subsets, and that $P_C(C')=\frac{P_D(C')}{P_D(C)}$, we obtain from (\ref{eq:homCC'})
\begin{equation*}
f(P_D(C'))=f(P_D(C))+f(P_D(C')/P_D(C)),
\end{equation*}
from which it follows using standard additivity arguments, monotonicity and continuity of the function $f(\cdot)$ (which follow from Postulates Q6 and Q2, respectively) that $f(x)=-c \log x$ where $c>0$ is a constant (see \cite{RENYI:1961} for details). Since the constant $c$ may depend on the particular homogeneous subset $D$ we can denote it by $u(D)$ and obtain that
\begin{equation}
G(D,C,P_D)=-u(D)\log P_D(C),
\label{eq:GhomD}
\end{equation}
for any $C\subset D$ whenever $D$ is homogeneous.

Substituting (\ref{eq:GhomD}) into (\ref{eq:homDC}) we can obtain
\begin{equation*}
G(\Omega,C,P)=G(\Omega,D,P)+f(P(C)/P(D))=G(\Omega,D,P)-u(D)\log \frac{P(C)}{P(D)},
\end{equation*}
or, equivalently,
\begin{equation}
G(\Omega,C,P)-G(\Omega,D,P)=-u(D)\log P(C) - u(D)\log P(D),
\label{eq:G(D)}
\end{equation}
where $C$ is an arbitrary subset of $D$. Then it follows from (\ref{eq:G(D)}) and continuity of the function $G$ (Postulate Q2) that
\begin{equation}
G(\Omega, C, P)= -u(D)\log P(C) + v(D),
\label{eq:GhomOmv}
\end{equation}
for any $C\subseteq D$ whenever $D$ is a homogeneous subset of $\Omega$. Here $v(D)$ is an arbitrary function of $D$. Setting $P(C)=1$ in (\ref{eq:GhomOmv}) and making use of Postulate Q1, we obtain that $v(D)\equiv 0$ and therefore
\begin{equation}
G(\Omega, C, P)= -u(D)\log P(C).
\label{eq:GhomOm}
\end{equation}

Now let $\bD=\{D_1,\dotsc, D_N\}$ be a complete partition of $\Omega$ into homogeneous subsets $D_j$, $j=1,\dotsc, N$. Let $C\subset \Omega$ be an incomplete ideal question. Then $C=\cup_{j=1}^N C\cap D_j$, and since $D_j$ is homogeneous and $C\cap D_j\subseteq D_j$, we obtain using (\ref{eq:GhomOm}) that
\begin{equation}
G(\Omega, C\cap D_j, P)= -u(D_j)\log P(C\cap D_j).
\label{eq:G(OmCcapDP)}
\end{equation}
On the other hand, by Postulate Q3,
\begin{equation}
G(\Omega, C, P) = G(\Omega, \bD_C, P)-G(C,\bD_C,P_C),
\label{eq:G(OmCP)}
\end{equation}
where
\begin{equation}
G(\Omega, \bD_C, P)=-\frac{1}{P_C} \sum_{j=1}^N u(D_j)P(C\cap D_j)\log P(C\cap D_j),
\label{eq:G(OmDP)}
\end{equation}
(using the identity $C=\cup_{j=1}^N C\cap D_j$, expression (\ref{eq:G(OmCcapDP)}) and Postulate Q4),
and analogously,
\begin{equation}
G(C, \bD_C, P_C)=-\sum_{j=1}^N u(D_j)\frac{P(C\cap D_j)}{P(C)}\log \frac{P(C\cap D_j)}{P(C)},
\label{eq:G(CDPC)}
\end{equation}

Substituting (\ref{eq:G(OmDP)}) and (\ref{eq:G(CDPC)}) into (\ref{eq:G(OmCP)}) we obtain
\begin{equation}
G(\Omega, C, P)=-\sum_{j=1}^N \frac{P(C\cap D_j)}{P(C)} u(D_j) \log P(C).
\label{eq:G(OmCPcomp)}
\end{equation}
We can rewrite (\ref{eq:G(OmCPcomp)}) as
\begin{equation}
G(\Omega, C, P)=-u(C)P(C) \log P(C),
\label{eq:G(OmCPres)}
\end{equation}
where
\begin{equation}
u(C)\equiv \sum_{j=1}^N \frac{P(C\cap D_j)u(D_j)}{P(C)}
\label{eq:u(C)def}
\end{equation}
can be thought of as the definition of function $u$: $\mF \rightarrow \bR$ for inhomogeneous subsets of $\Omega$. If we define the function $u(\o)$ on $\Omega$ by
\begin{equation*}
u(\o)=\sum_{j=1}^N u(D_j) I_{D_j}(\o),
\end{equation*}
where $I_D(\o)$ is the indicator function of a subset $D\subseteq \Omega$, then the expression (\ref{eq:u(C)def}) can be written as
\begin{equation}
u(C)=\frac{\int_C u(\o) dP(\o)}{P(C)}.
\label{eq:u(C)}
\end{equation}
Finally, if $\bC=\{C_1,\dotsc, C_r\}$ is an arbitrary question, we can use (\ref{eq:G(OmCPres)}) and Postulate Q4 to obtain
\begin{equation*}
G(\Omega,\bC,P)=\frac{-\sum_{j=1}^r u(C_j)P(C_j)\log P(C_j)}{\sum_{j=1}^r P(C_j)},
\end{equation*}
where the ``weights'' $u(C_j)$ of the subsets $C_j$ are given by (\ref{eq:u(C)}). \qed

Theorem~\ref{th:G} establishes the general form of the question difficulty functional if isotropy and linearity conditions are imposed. The result depends on the measure $P$ and an integrable function $u(\cdot)$ on the parameter space $\Omega$ that can be thought of as a description of the information source's knowledge structure. Note that while the measure is extensive, i.e. the measure of a union of two disjoint subsets of $\Omega$ is the sum of individual measures ($P(C\cup C')=P(C)+P(C')$ if $C\cap C'= \emptyset$), the function $u$ represents an intensive quantity in that it averages for a union of two disjoint subsets ($u(C\cup C')=\frac{P(C)u(C)+P(C')u(C')}{P(C)+P(C')}$). One can say, loosely speaking, that while measure is similar to volume, $u$ is similar to temperature if parallels with physics are to be used. These parallels suggest that the function $u(\o)$ can be thought of as temperature-like quantity that is allowed to be different at different points of the parameter space. In the following, we refer to the function $u(\o)$ as {\it intensity} or {\it pseudotemperature}. For the same reason, as mentioned earlier, it is convenient to think of question difficulty as the amount of {\it pseudoenergy} associated with the question.

It is also convenient to introduce the notion of {\it entropy} of question $\bC$ in the usual way as Shannon entropy of the probability distribution induced by the partition $\bC$:
$$H(\Omega,\bC,P)=\frac{\sum_{j=1}^r P(C_j)\log \frac{1}{P(C_j)}}{\sum_{j=1}^r P(C_j)},$$
which differs from the pseudoenergy (difficulty) in that it does not involve the pseudotemperature $u(\cdot)$. It is easy to see that, for an ideal question $C\subset \Omega$, the relationship between pseudoenergy and entropy is simply
$$G(\Omega, C, P)=u(C)H(\Omega,C,P), $$
that is identical to the relationship that exists between thermal energy (heat) and entropy in thermodynamics for reversible processes.

The form of the question difficulty functional, as has been just mentioned, is determined up to an arbitrary non-negative integrable function on the base space $\Omega$. This function has the meaning of a description of the knowledge structure of a given information source and therefore is in general different for different information sources. If an agent is faced with an ``unfamiliar'' information source, he or she would in principle need to estimate the source's knowledge structure expressed by its pseudotemperature function. This function would describe the source's strengths and weaknesses with regards to its ability of producing accurate answers to various questions. The only practical way of estimating the source's pseudotemperature function $u(\cdot)$ is by ``probing'' the source's knowledge \footnote{Note that this is indeed the method that has been in use for a long time in many practical applications only without being formalized in any regular fashion. For example, this is what good teachers in pretty much all areas do to figure out their students' knowledge state so that an effective course of further study (typically, concentrated on working on the weaknesses) can be developed.} asking the source some ``sample'' questions and comparing the answers to actual outcomes once the latter become available. The specific methods for estimating the function $u(\cdot)$ are discussed in the companion paper \cite{part2}.

Observe also that, if one multiplied the function $u(\o)$ by any positive constant, the difficulty of {\it any} question would be multiplied by the same constant. This means $u(\o)$ is defined up to an overall scale that is equivalent to a choice of pseudoenergy measuring units. This overall scale can be thus chosen according to a convention of the agent's choice. One such convention that we will use is that the pseudotemperature is normalized so that its average value on the whole base space is unity:
\begin{equation*}
\int_{\Omega} u(\o) dP(\o)=1.
\end{equation*}
One useful property of this convention is that, if the pseudotemperature is constant on $\Omega$, i.e. the whole $\Omega$ is homogeneous, pseudoenergy coincides with entropy and can be measured in the familiar binary ``bits''. Another useful convention is described in the companion paper \cite{part2}.

 \section{\label{s:relations}Relationships between different questions}
 In this section, we assume that all questions are complete.
 If $\bC'$ and $\bC''$ are two arbitrary (complete) questions, the expression $\sum_{C'\in \bC'} P(C')G(C',\bC''_{C'},P_{C'})$ will be denoted $G(\Omega,\bC''_{\bC'},P)$ and called the {\it conditional difficulty}, or, equivalently, {\it conditional pseudoenergy} of $\bC''$. Using this notation, the sequentiality property expressed by Postulate Q3$'$ can be rewritten as
\begin{equation}
G(\Omega,\tilde\bC,P)= G(\Omega,\bC,P)+G(\Omega,\tilde\bC_{\bC},P),
\label{eq:seq_c}
\end{equation}
where $\tilde\bC$ is an arbitrary refinement of $\bC$.

If $\bC'$ and $\bC''$ are two arbitrary (complete) questions and $\bC=\bC'\cap \bC''$ then obviously $\bC$ is a refinement of both $\bC'$ and $\bC''$. One can then write the sequentiality property (\ref{eq:seq_c}) as
\begin{equation}
G(\Omega, \bC, P)=G(\Omega, \bC', P)+G(\Omega, \bC_{\bC'},P).
\label{eq:seq_c1}
\end{equation}
But it is easy to see that the partition induced by $\bC=\bC'\cap \bC''$ on any set $C'$ in $\bC'$ is exactly the same as the partition induced on that set by $\bC''$. Therefore, the term $G(\Omega, \bC_{\bC'},P)$ in (\ref{eq:seq_c1}) can be equivalently written as $G(\Omega, \bC''_{\bC'},P)$ and we arrive at the {\it chain rule} for the question difficulty which we formulate as a lemma.

\begin{lemma}
If $\bC'$ and $\bC''$ are two arbitrary complete questions and $P$ is a measure on $\Omega$ then
$$G(\Omega, \bC'\cap \bC'', P)=G(\Omega, \bC', P)+G(\Omega, \bC''_{\bC'},P). $$
\end{lemma}

\begin{figure}
\includegraphics[scale=0.6]{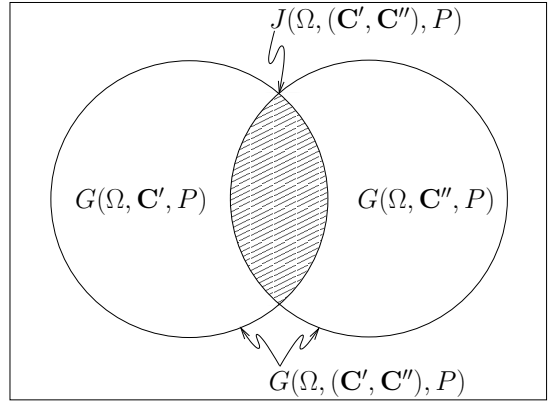}
\caption{\label{f:overlap}Venn diagram for pseudoenergy overlap.}
\end{figure}

Again, let $\bC'$ and $\bC''$ be two (complete) questions on $\Omega$ and let $\bC=\bC'\cap \bC''$ be the resulting combined question. Then the {\it pseudoenergy overlap} $J(\Omega,(\bC';\bC''),P)$ between $\bC'$ and $\bC''$ can be defined as the difference between the sum of difficulties of $\bC'$ and $\bC''$ and that of the combined question $\bC'\cap \bC''$:
\begin{equation}
J(\Omega,(\bC';\bC''),P)=G(\Omega,\bC',P)+G(\Omega,\bC'',P)-G(\Omega,\bC'\cap \bC'',P)
\label{eq:overlap}
\end{equation}
The definition (\ref{eq:overlap}) can be illustrated by a Venn diagram (see~Fig.~\ref{f:overlap}). Note that $J(\Omega,(\bC';\bC''),P)$ is symmetric with respect to $\bC'$ and $\bC''$. From the point of view of the distributive lattice of questions, $\bC=\bC'\cap \bC''$ is the question representing the meet of questions $\bC'$ and $\bC''$. The join of these questions will in general be not a partition question but instead a question that can be associated with a collection of overlapping subsets of $\Omega$. The pseudoenergy overlap $J(\Omega,(\bC';\bC''),P)$ can then be identified with the difficulty of the join of $\bC'$ and $\bC''$. The relation (\ref{eq:overlap}) then becomes nothing else but the {\it sum rule} for valuations and bi-valuations on lattices.

One can make use of the sequentiality property of pseudoenergy to rewrite expression for the pseudoenergy overlap as follows.
\begin{align*}
J(\Omega,(\bC';\bC''),P) &= G(\Omega,\bC',P)+G(\Omega,\bC'',P)-G(\Omega,(\bC',\bC''),P)\\
 &= G(\Omega,\bC',P)+G(\Omega,\bC'',P) - G(\Omega,\bC'',P) - G(\Omega,\bC'_{\bC''},P)\\
&= G(\Omega,\bC',P)- G(\Omega,\bC'_{\bC''},P).
\end{align*}
We formulate this result as a lemma.

\begin{lemma}
If $\bC'$ and $\bC''$ are two arbitrary questions and $P$ is a measure on $\Omega$ then the pseudoenergy overlap $J(\Omega,(\bC';\bC''),P)$ can be found as
$$J(\Omega,(\bC';\bC''),P)=G(\Omega,\bC',P)- G(\Omega,\bC'_{\bC''},P).$$
\label{l:overlap}
\end{lemma}

Clearly, due to symmetry, the expression for the pseudoenergy overlap stated in Lemma~\ref{l:overlap} can be equivalently written as $J(\Omega,(\bC';\bC''),P)=G(\Omega,\bC'',P)- G(\Omega,\bC''_{\bC'},P) $.

If an expression for the pseudoenergy overlap as a function of the measure $P$ and the pseudotemperature  $u(\o)$ is desired the definition (\ref{eq:overlap}) together with Theorem~\ref{th:G} can be used to obtain
\begin{equation}
J(\Omega,(\bC';\bC''),P) = \sum_{i=1}^{r'} \sum_{j=1}^{r''} u(C'_i\cap C''_j)P(C'_i\cap C''_j)
\log \frac{P(C'_i\cap C''_j)}{P(C'_i) P(C''_j)}.
\label{eq:overlap(uP)}
\end{equation}

We will be interested in exploring relationships between different questions: given two distinct questions, we would like to know to what degree they are similar to each other. More specifically, if a perfect answer to one question is available, how the difficulty of the other question is affected. To answer this question, let $\bC'$ and $\bC''$ be two arbitrary complete questions on $\Omega$ and let $V^*(\bC')$ be a perfect answer to $\bC'$. We would like to find an expression for the conditional difficulty of $\bC''$ given $V^*(\bC')$. Clearly, since a reception of value $s'_j$ of $V(\bC')$ updates the measure $P$ to $P_{C'_j}$, the difficulty of $\bC''$ given $V(C')=s'_j$ is equal to
\begin{equation}
G(\Omega, \bC'', P_{C'_j}) = G(C'_j, \bC''_{C'_j}, P_{C'_j}),
\label{eq:Gcond}
\end{equation}
since subsets of zero measure do not contribute to the difficulty. Therefore the
overall (expected) difficulty $G(\Omega,\bC'',V^*(\bC'))$ of question $\bC''$ given a perfect answer $V^*(\bC')$ to $\bC'$ can be written as
\begin{equation}
\label{eq:GC''V'}
\begin{split}
 G(\Omega,\bC'',V^*(\bC')) &= \sum_{j=1}^{r'} \Pr(V^*(\bC')=s_j) G(\Omega, \bC'', P_{C'_j})\\
              &\overset{(a)}{=} \sum_{j=1}^{r'} P(C'_j) G(C'_j, \bC''_{C'_j}, P_{C'_j})= G(\Omega, \bC''_{\bC'}, P)\\
                &\overset{(b)}{=} G(\Omega, \bC'',P)-J(\Omega, (\bC';\bC''), P),
\end{split}
\end{equation}
where (a) follows from (\ref{eq:Gcond}) and the consistency condition (\ref{eq:cons-mc}) -- which implies that $\Pr(V^*(\bC')=s_j)=P(C_j)$; (b) follows from Lemma~\ref{l:overlap}.

We see from (\ref{eq:GC''V'}) that the conditional difficulty of $\bC''$ can be represented as a difference of the standard (unconditional) difficulty and the pseudoenergy overlap $J(\Omega, (\bC';\bC''), P)$. Thus the latter provides a measure of reduction of difficulty of a question that is due to a perfect knowledge of an answer to another question. Such a measure can naturally be termed {\it relative depth} \footnote{Strictly speaking, the notion of {\it answer depth} is defined and discussed only in the companion paper \cite{part2}, so it is used here a little ``ahead of time''.} of an answer $V(\bC')$ (which in general may not be perfect) with respect to question $\bC''$. We can formulate the result just obtained as a lemma.

\begin{lemma}
The relative depth of a perfect answer $V^*(\bC')$ to question $\bC'$ with respect to question $\bC''$ is equal to the pseudoenergy overlap between questions $\bC'$ and $\bC''$.
\label{l:rel-inf}
\end{lemma}

The result of Lemma~\ref{l:rel-inf} has a clear intuitive interpretation: If two distinct questions are close, i.e. ``almost about the same thing'' then knowing a (perfect) answer to one of them nearly answers the other one -- reduces the difficulty of it to a small value compared to the initial difficulty. The pseudoenergy overlap quantifies the notion of closeness for two arbitrary questions.

\section{\label{s:examples}Examples}
We consider an example with a finite parameter space first. Let $\Omega$ consist of 8 elements, corresponding to green, yellow and red apples (denoted $GA$, $YA$ and $RA$, respectively), green, yellow and red pears (denoted $GPr$, $YPr$ and $RPr$), and yellow and red peaches (denoted $YPc$ and $RPc$). Let all elements be equiprobable so that $P(\cdot)=\frac{1}{8}$ for all $\o\in \Omega$. The function $u(\o)$ describes the relative difficulty of respective ideal questions. To this effect, let us suppose that it was found (say, using estimation procedures described in \cite{part2}) that $u(GA)=u(GPr)=1$, $u(YPr)=u(RPr)=1.5$ and $u(YA)=u(RA)=u(YPc)=u(RPc)=2$. Normalizing the values of $u(\cdot)$ so that $\int_{\Omega} u(\o) dP(\o)=1$ one obtains $u(GA)=u(GPr)=\frac{8}{13}$, $u(YPr)=u(RPr)=\frac{12}{13}$ and $u(YA)=u(RA)=u(YPc)=u(RPc)=\frac{16}{13}$.

The difficulties of ideal questions corresponding to individual elements of $\Omega$ can be found as follows: $G(\Omega, GA, P)=G(\Omega, GPr, P)=\frac{8}{13}\cdot \log 8=\frac{24}{13}$, $G(\Omega, YPr, P)=G(\Omega, RPr, P)=\frac{12}{13}\cdot \log 8=\frac{36}{13}$ and $G(\Omega, YA, P)=G(\Omega, RA, P)=G(\Omega, YPc, P)=G(\Omega, RPc, P)=\frac{16}{13}\cdot \log 8=\frac{48}{13}$. The difficulty of the exhaustive question (that asks to determine the type and color of the fruit presented to the source) can be found as an expectation of the difficulties of all these ideal questions. Denoting the corresponding (finest) partition of $\Omega$ by $\bC_f$ we obtain
$$ G(\Omega, \bC_f, P)=\sum_{\o\in \Omega} P(\o)G(\Omega, \o, P)=3.$$
Now let us consider difficulties of other complete questions. Let first of such questions be {\it ``Is the fruit green or not?''}. Let $C_g=\{GA, GPr\}\subset \Omega$ be the subset consisting of all green fruit (apples and pears) and let $\overline C_g=\Omega\setminus C_g$ be the subset containing fruit of all other colors (red and yellow). The values $u(\cdot)$ for the sets in this partition are $u(C_g)=\frac{8}{13}$ and $u(\overline C_g)=\frac{1}{3}\cdot \frac{12}{13}+\frac{2}{3}\cdot \frac{16}{13}=\frac{44}{39}$. The measures are $P(C_g)=\frac{1}{4}$ and $P(\overline C_g)=\frac{3}{4}$.
Thus the difficulty of the question {\it ``Is the fruit green or not?''} can be found as
$$ G(\Omega, \{C_g, \overline C_g\}, P)=u(C_g)P(C_g)\log \frac{1}{P(C_g)}+u(\overline C_g)P(\overline C_g)\log \frac{1}{P(\overline C_g)}=0.66 $$

Consider another question with subset measures (and thus ``metric elaborateness'') equal to those of $\{C_g, \overline C_g\}$. The question is {\it ``Is the fruit a peach or not?''}. The corresponding partition is $\{C_{Pc}, \overline C_{Pc}\}$ where $C_{Pc}=\{YPc, RPc\}$ and $\overline C_{Pc}=\Omega\setminus C_{Pc}$. The values of function $u(\cdot)$ on these subsets are $u(C_{Pc})=\frac{16}{13}$ and $u(\overline C_{Pc})=\frac{1}{3}\cdot \frac{8}{13}+\frac{1}{3}\cdot \frac{12}{13}+\frac{1}{3}\cdot \frac{16}{13}=\frac{12}{13}$. The measures are $P(C_{Pc})=\frac{1}{4}$ and  $P(\overline C_{Pc})=\frac{3}{4}$. The difficulty of the question $\{C_{Pc}, \overline C_{Pc}\}$ is
$$G(\Omega, \{C_{Pc}, \overline C_{Pc}\}, P)= u(C_{Pc})P(C_{Pc})\log \frac{1}{P(C_{Pc})}+u(\overline C_{Pc})P(\overline C_{Pc})\log \frac{1}{P(\overline C_{Pc})}= 0.90$$

We see that this question is somewhat more difficult than the question on whether the fruit is green. The main reason for this difference is that to answer the question on whether the fruit is a peach one might need to have to tell a peach from an apple of similar (warm) color which is relatively difficult while answering the question on whether the fruit is green does not involve any ``hard'' decisions since the color itself is distinct.

Consider now the question {\it ``What color is the given fruit?''} on one hand and {\it ``What type is the given fruit?''} on the other. The former question can be represented as the partition $\bC_c=\{C_g, C_y, C_r \}$ where $C_g=\{GA, GPr\}$, $C_y=\{YA, YPr, YPc\}$ and $C_r=\{RA, RPr, RPc\}$; the latter question can be identified with the partition $\bC_t=\{C_A, C_{Pr}, C_{Pc} \}$ where $C_A=\{GA, YA, RA\}$, $C_{Pr}=\{GPr, YPr, RPr\}$ and $C_{Pc}=\{YPc, RPc\}$. The values of $u(\cdot)$ on these subsets are $u(C_g)=\frac{8}{13}$, $u(C_y)=\frac{1}{3}\cdot \frac{12}{13}+\frac{2}{3}\cdot \frac{16}{13}=\frac{44}{39}$, $u(C_g)=u(C_y)=\frac{44}{39}$; $u(C_A)=\frac{1}{3}\cdot \frac{8}{13}+\frac{2}{3}\cdot \frac{16}{13}=\frac{40}{39}$, $u(C_{Pr})=\frac{1}{3}\cdot \frac{8}{13}+\frac{2}{3}\cdot \frac{12}{13}=\frac{32}{39}$, $u(C_{Pc})=\frac{16}{13}$. The measures are $P(C_g)=\frac{1}{4}$, $P(C_y)=\frac{3}{8}$, $P(C_r)=\frac{3}{8}$; $P(C_A)=P(C_{Pr})=\frac{3}{8}$, $P(C_{Pc})=\frac{1}{4}$. Thus the difficulties of these two questions are
\begin{align*}
G(\Omega, \{C_g, C_y, C_r\}, P)&= u(C_{g})P(C_{g})\log \frac{1}{P(C_{g})}+u(C_{y})P(C_{y})\log \frac{1}{P(C_{y})}\\ &+u(C_{r})P(C_{r})\log \frac{1}{P(C_{r})}=\frac{11}{13}\log \frac{8}{3}+\frac{2}{13}\log 4 =1.51,
\end{align*}
and
\begin{align*}
G(\Omega, \{C_A, C_{Pr}, C_{Pc}\}, P)&= u(C_{A})P(C_{A})\log \frac{1}{P(C_{A})}+u(C_{Pr})P(C_{Pr})\log \frac{1}{P(C_{Pr})}\\ &+u(C_{Pc})P(C_{Pc})\log \frac{1}{P(C_{Pc})}=\frac{9}{13}\log \frac{8}{3}+\frac{4}{13}\log 4 = 1.60,
\end{align*}
respectively.

The question about color turns out to be slightly easier than that about type. Qualitatively, the main reason for this difference is that the relatively rare event (that the fruit is green and that it is a peach, respectively) that gives a larger contribution to the difficulty because of the $\log \frac{1}{P(\cdot)}$ factor has smaller average value of pseudotemperature $u(\cdot)$ in the case of the question about the fruit color.

The pseudoenergy overlap between the ``color'' and ``type'' questions can be calculated using the expression (\ref{eq:overlap(uP)}):
$$J(\Omega, (\bC_c; \bC_t), P)=\frac{6}{13}\log \frac{4}{3}+\frac{7}{13}\log \frac{8}{9}=0.100, $$
indicating that while a perfect knowledge of the fruit color helps answering the question about its type, the reduction of difficulty of the ``type'' question due to the knowledge of color is relatively mild so the question about the fruit type remains almost as hard as it was before the color became known.

\begin{figure}
\includegraphics[scale=0.7]{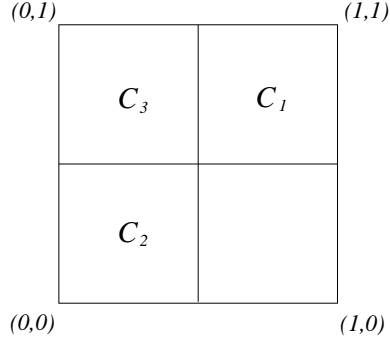}
\caption{\label{f:questions-square}The parameter space $\Omega=[0,1]^2$ and subsets $C_i$, $i=1,2,3$.}
\end{figure}

For an example with infinite parameter space, consider $\Omega=[0,1]^2$ with uniform measure $P$ (see Fig.~\ref{f:questions-square} for an illustration). Let $u(\o)=\frac{3}{2}(\o_1^2+\o_2^2)$ where $\o_1$ and $\o_2$ are coordinates on $\Omega$. Let us consider three different questions: $\bC_i=\{C_i,\overline C_i\}$, where $C_1=\{\o: \o_1\in [\frac{1}{2},1], \o_2\in [\frac{1}{2},1] \}$, $C_2=\{\o: \o_1\in [0,\frac{1}{2}], \o_2\in [0,\frac{1}{2}] \}$, $C_3=\{\o: \o_1\in [0,\frac{1}{2}], \o_2\in [\frac{1}{2},1] \}$. It is easy to see that $P(C_i)=\frac{1}{4}$ for $i=1,2,3$.

For question $\bC_1$, we have $u(C_1)=\frac{3}{2}\int_{\frac{1}{2}}^1 d\o_1 \int_{\frac{1}{2}}^1 d\o_2(\o_1^2+\o_2^2) =\frac{7}{4}$. Then, using the normalization condition $u(C_1)P(C_1)+u(\overline C_1)P(\overline C_1)=1$, we can obtain $u(\overline C_1)= \frac{3}{4}$, which allows us to compute the difficulty:
$$ G(\Omega, \{C_1,\overline C_1\}, P)= u(C_1)P(C_1)\log \frac{1}{P(C_1)}+u(\overline C_1)P(\overline C_1)\log \frac{1}{P(\overline C_1)}=\frac{1}{2}\log \frac{4}{3}+\frac{1}{2}\log 4=1.208.$$

For question $\bC_2$, we obtain $u(C_2)=\frac{3}{2}\int_0^{\frac{1}{2}} d\o_1 \int_0^{\frac{1}{2}} d\o_2(\o_1^2+\o_2^2) =\frac{1}{16}$, and, making use of the normalization condition, $u(\overline C_2)= \frac{21}{16}$. The difficulty functional value for this question becomes
$$ G(\Omega, \{C_2,\overline C_2\}, P)= u(C_2)P(C_2)\log \frac{1}{P(C_2)}+u(\overline C_2)P(\overline C_2)\log \frac{1}{P(\overline C_2)}=\log \frac{4}{3}=0.415.$$

Finally, for question $\bC_3$, we have $u(C_3)=\frac{3}{2}\int_0^{\frac{1}{2}} d\o_1 \int_{\frac{1}{2}}^1 d\o_2(\o_1^2+\o_2^2) =1$, and, obviously, $u(\overline C_3)= 1$. The difficulty functional is
$$ G(\Omega, \{C_3,\overline C_3\}, P)= u(C_3)P(C_3)\log \frac{1}{P(C_3)}+u(\overline C_3)P(\overline C_3)\log \frac{1}{P(\overline C_3)}=\frac{3}{4}\log \frac{4}{3}+\frac{1}{4}\log 4=0.811.$$

We see that, among these three questions $\bC_1$ turns out to be the most difficult while difficulty of $\bC_2$  is the smallest of the three. The reason is that $\bC_1$ includes a small measure (rare) set in the region of high values of pseudotemperature $u(\o)$. On the other hand, the rare subset in $\bC_2$ is located in the region of small values of $u(\o)$. Question $\bC_3$ is naturally placed between these two extremes: its rare subset is located in the region of moderate values of the field $u(\o)$ so that the difficulty weight of this subset is equal to the average for the whole parameter space.

The overlaps between these questions can easily be computed using expression (\ref{eq:overlap(uP)}).
$$J(\Omega, (\bC_1;\bC_2),P)=\frac{1}{2}\log \frac{4}{3}+ \frac{1}{2}\log \frac{8}{9}=0.123,$$
$$J(\Omega, (\bC_1;\bC_3),P)=\frac{11}{16}\log \frac{4}{3}+ \frac{5}{16}\log \frac{8}{9}=0.232,$$
and
$$J(\Omega, (\bC_2;\bC_3),P)=\frac{5}{16}\log \frac{4}{3}+ \frac{11}{16}\log \frac{8}{9}=0.013,$$
showing that the most difficult questions -- $\bC_1$ and $\bC_3$ -- also exhibit the largest overlap which agrees with the common sense derived notion that a knowledge of a perfect answer to a more difficult question can give more help in answering another question.

It is interesting to consider the limit in which the measure of the rare set approaches zero. For this purpose, let $C_1=\{\o: \o_1\in [1-a,1], \o_2\in [1-a,1] \}$, $C_2=\{\o: \o_1\in [0,a], \o_2\in [0,a] \}$ and $C_3=\{\o: \o_1\in [0,a], \o_2\in [1-a,1] \}$ and let $\bC_i=\{C_i, \overline C_i\}$ for $i=1,2,3$.
Let $u(\o)=\frac{n+1}{2}(\o_1^n+\o_2^n)$ where $n\ge 2$ is an integer and $\o\in \Omega=[0,1]^2$. Then repeating the calculations for the previously considered example, taking the limit $a\rightarrow \infty$ and retaining only terms of the lowest order in $a$ we obtain
$$G(\Omega, \{C_1,\overline C_1\}, P)\simeq (n+1) a^2\log \frac{1}{a}+\log e\cdot a^2\simeq (n+1) a^2\log \frac{1}{a},$$
$$G(\Omega, \{C_2,\overline C_2\}, P)\simeq \log e\cdot a^2,$$
and
$$G(\Omega, \{C_3,\overline C_3\}, P)\simeq 2 a^2\log \frac{1}{a}+\log e\cdot a^2\simeq 2a^2\log \frac{1}{a}.$$
Again, we can see that the question $\bC_1$ ends up being the most difficult one, with $\bC_2$ being the least difficult. It's interesting to note that, to leading order in $a$, the difficulty of $\bC_1$ and $\bC_3$ behaves as $a^2\log \frac{1}{a}$ (with only a numerical coefficient being different), while the difficulty of $\bC_2$ behaves as $a^2$. A related observation is that, in this limit, the difficulty of both $\bC_1$ and $\bC_3$ is dominated by the rare subset while that of $\bC_2$ is dominated by the larger subset with measure approaching 1 since the contribution of the rare subset is diminished by the low value of pseudotemperature $u(\cdot)$ over that subset.

\section{\label{s:conclusion}Conclusion}
This article initiated development of a general quantitative framework for the description of the process of information extraction from information sources capable of providing answers to given questions. The main motivation for such a framework is the need for optimal decision making in situations characterized with incomplete information and availability of additional information sources. The framework is expected to be especially useful when the knowledge the information sources possess is of a relatively ``loose'' variety, i.e. cannot be readily represented in a form admitting direct use in a mathematical formulations. A typical example of such a source would be a human expert who can express a preference for one of the two regions in the parameter space but would find it difficult to produce an accurate probability distribution over the parameter space.

The main components of the proposed framework are questions, answers and information sources. The present article's subject is questions and, in particular, question difficulty functionals. The purpose of the latter is measuring the degree of expected accuracy the given source can achieve answering various questions. The idea is that a source would answer easy question well but its answers' accuracy would decrease with increasing difficulty of questions. The overall form of the question difficulty functional is in general determined by the constraints the difficulty functional is required to satisfy.  In this article we made an assumption that the question difficulty is linear and isotropic on the parameter space. The resulting form was then derived from a system of postulates expressing the desired properties along with more general consistency requirements.

It turns out that the resulting question difficulty functional depends on a single scalar quantity $u(\cdot)$ defined on the parameter space and can be interpreted -- using parallels with thermodynamics -- as an energy-like quantity while the function $u(\cdot)$ takes on the role of temperature that is allowed to take different values at different points of the parameter space. It is interesting to contrast the resulting difficulty functional to the corresponding Shannon entropy that can be interpreted as a quantity measuring the minimum expected number of bits required to communicate a (perfect) answer to the question under consideration. Using parallels with thermodynamics, one can say that, while the former is akin to thermal energy, the latter can be likened to entropy.

It is worth noting, as was alluded to in Section~\ref{s:intro}, that the system of postulates used in the present article is somewhat restrictive in that it has the isotropy of the source's knowledge structure ``built in''. For instance, the proposed system of postulates implicitly assumes that an ideal question difficulty is a well-defined quantity, regardless of the real question it is ``an aspect of'' (in the sense explained in Section~\ref{s:frame}). The consequence of that is the resulting source knowledge structure -- and hence the form of the difficulty functional -- is described by a {\it scalar} function on the base space. On the other hand, imagine, for example, an information source that can answer questions about fruit color a lot better than those about its kind. In this case, the source's knowledge structure would exhibit a pronounced {\it directional} dependence. Preliminary investigations show that such knowledge structures can be captured if the system of postulates is weakened somewhat to exclude implicit assumptions mentioned above. In such cases, the difficulty functional can be shown to depend on an arbitrary rank-2 {\it tensor} as opposed to a scalar function in the isotropic case described in the present article. Details will appear in future publications.

Note also that while we used a ``traditional style'' axiomatic approach to determine the form of the question difficulty functional, we believe that it can be derived along the lines of the more recent  order-theoretic approach described in \cite{knuth05,knuth08} as was already mentioned in the Introduction. Specifically, the question relevance measure introduced in \cite{knuth05} is a natural bi-valuation on the distributive lattice of questions defined as down-sets of all subsets of the elements of the lattice of logical assertions. The partial order on the lattice of questions is then given by set inclusion which can be interpreted as ``answering''. The relevance (which, we believe, would be better called ``bearing'' as originally suggested by Cox) was defined in \cite{knuth05,knuth08} as a bi-valuation on the sublattice of real questions that gives the degree to which a real question resolves the central issue. Since partition (complete partition in our terminology) questions are the join-irreducible elements of the real lattice,
their valuations can be assigned arbitrarily \cite{rota1971}. Suppose now one wants the question valuations to be related to the valuations on the corresponding lattice of logical assertions.  Then if one makes a single assumption that the valuation for a partition question depends only on probabilities of assertions (subsets of $\Omega$ in our interpretation) constituting this question, then the constraints imposed by the lattice structure lead to the unique form of the valuation equal (up to one multiplicative and one additive constant) to the Shannon entropy. On the other hand, if one instead assumes that the question valuation depends -- besides  the subset probabilities that encode the initial information about the system -- on some geometric object on the base space $\Omega$ that describes the particular information source's knowledge structure, we believe that one would recover the difficulty functional. The nature of the geometric object would then be determined by symmetry considerations. This issue is currently under investigation.




%

\end{document}